\documentclass[aps,onecolumn,preprintnumbers,pra]{revtex4-2}

\usepackage{graphicx}  
\usepackage{subfigure}
\usepackage{multirow}
\usepackage{mathtools}
\usepackage{notes2bib}
\usepackage{nicefrac,xfrac}
\Large

\usepackage[utf8]{inputenc}
\usepackage[english]{babel}

\usepackage{fancyhdr}
\usepackage{longtable}
\usepackage{parskip}
\usepackage[T1]{fontenc}
\usepackage{dcolumn}   

\usepackage{bm}        
\usepackage{amsfonts}  
\usepackage{amsmath}   
\usepackage{amssymb}   

\newcommand{\pwisein}{\left\{ \begin{array}{ll}}
\newcommand{\pwiseout}{\end{array}\right.}

\usepackage[unicode=true,pdfusetitle,
 bookmarks=true,bookmarksnumbered=false,bookmarksopen=false,
 breaklinks=false,pdfborder={0 0 0},pdfborderstyle={},backref=false,colorlinks=true]
 {hyperref}
\hypersetup{allcolors=blue}

\makeatletter
\renewcommand\NAT@citesuper[3]{\ifNAT@swa
\unskip\hspace{1\p@}\textsuperscript{[#1]}%
\if\relax#3\relax\else\ [#3]\fi\else [#1]\fi\endgroup}
\makeatother

\usepackage [english]{babel}
\usepackage [autostyle, english = american]{csquotes}
\MakeOuterQuote{"}

\begin{document}
\setcitestyle{super}
\title{The robustness of composite pulses elucidated by classical mechanics:\linebreak Stability around the globe}

\author{Jonathan Berkheim}
\email{jonathan.berkheim@weizmann.ac.il}
\author{David J. Tannor}

\affiliation {\it Department of Chemical and Biological Physics, Weizmann Institute of Science, 76100, Rehovot, Israel}


\begin{abstract}  
Composite Pulses (CPs) are widely used in Nuclear Magnetic Resonance (NMR), optical spectroscopy, optimal control experiments and quantum computing to manipulate systems that are well-described by a two-level Hamiltonian. A careful design of these pulses can allow the refocusing of an ensemble at a desired state, even if the ensemble experiences imperfections in the magnitude of the external field or resonance offsets. Since the introduction of CPs, several theoretical justifications for their robustness have been suggested. In this work, we suggest another justification based on the classical mechanical concept of a stability matrix. The motion on the Bloch Sphere is mapped to a canonical system of coordinates and the focusing of an ensemble corresponds to caustics, or the vanishing of an appropriate stability matrix element in the canonical coordinates. Our approach highlights the directionality of the refocusing of the ensemble on the Bloch Sphere, revealing how different ensembles refocus along different directions. The approach also clarifies when CPs can induce a change in the width of the ensemble as opposed to simply a rotation of the axes. As a case study, we investigate the $90(x)180(y)90(x)$ CP introduced by Levitt, where the approach provides a new perspective into why this CP is effective.
\end{abstract}

\maketitle 
\section{Introduction: from a purely quantum problem to a purely classical treatment}
\subsection{Main goals}
Many experiments in atomic, molecular, and optical physics aim to invert a population of particles subjected to an external pulse. For example, in Nuclear Magnetic Resonance (NMR) an ensemble of nuclei that evolves under a constant field and a radiofrequency (RF) field has to undergo a synchronized population inversion between the two spin states in order to detect a considerable signal. In several fields of modern optics (e.g., pump-probe spectroscopy and quantum computing), it is sometimes desirable to invert an ensemble of electrons, initially localized in the ground state, to an excited electronic state in order to maximize the radiative signal or to perform other operations.

Consider an idealized ensemble of spins-1/2 in state $|a\rangle$ subjected to an external RF field. If one introduces simplifying assumptions such that (1) there are no correlations or couplings between spins, (2) the magnitude of the field is homogeneous along the ensemble and (3) the external RF field is on-resonance with the induced level separation, a monochromatic continuous-wave (CW) RF field will invert all spins to state $|b\rangle$ in half of the period associated with the Rabi frequency. However, when the ensemble experiences either an inhomogeneity in the magnitude of the field or a distribution of resonance offsets, the final state is a superposition of $|a\rangle$ and $|b\rangle$. The challenge is to find simple sequences of pulses that can produce population inversion under these circumstances for the entire ensemble.

It is with this challenge in mind that Composite Pulses (CPs) were invented. CPs consist of a short sequence of CW pulses, where each segment in the sequence has a characteristic amplitude, duration and phase. These pulses can compensate for the two imperfections described above: (1) field inhomogeneity and (2) resonance offset. Levitt and Freeman were the first to introduce such a pulse sequence,\citep{Levitt1979NMRPulse} though it was inspired by earlier notions, particularly, Hahn's Spin Echo\citep{Hahn1950SpinEchoes} and the Carr-Purcell-Meiboom-Gill sequence\citep{Carr1954EffectsExperiments,Meiboom1958ModifiedTimes} Since Levitt's original work, many extensions have been suggested.\citep{Glaser2015TrainingControl}

The first justification for the success of a CP was provided by Levitt himself, who used perturbation theory to develop an expression for the error caused by the pulse imperfections;\citep{Levitt1982SymmetricalOffset,Levitt1982SymmetricalOffsetb} later he developed a figure of merit based on the quaternion formalism of Blümlich and Spiess.\citep{Counsell1985AnalyticalPulses,Blumich1985QuaternionsRotations} An alternative approach based on Optimal Control theory was proposed by Boscain and Sugny.\citep{Glaser2015TrainingControl,Dridi2020RobustPulses} Each of these justifications can serve as a tool to design optimal pulses.

In this work, we suggest another justification based on the classical mechanical concept of a stability matrix. The motion on the Bloch Sphere is mapped to a canonical system of coordinates and the focusing of an ensemble corresponds to caustics, or the vanishing of an appropriate stability matrix element in the canonical coordinates. The caustics in CPs are unusual in the sense that each member is governed by a slightly different Hamiltonian, such that averaging is required for the stability analysis. Our approach highlights the directionality of the refocusing of the ensemble on the Bloch Sphere, revealing how different ensembles refocus along different directions. As a case study, we investigate the $90(x)180(y)90(x)$ CP introduced by Levitt, where the approach provides a new perspective into why this CP is effective: the focusing produced by Levitt's CP corresponds to a caustic, as manifested in the elements of the stability matrix. Levitt's perturbative treatment is seen to correspond to one element of the classical stability matrix. The approach clarifies why the $90(x)180(y)90(x)$ CP changes the width of the ensemble in the case of field inhomogeneity, as opposed to simply a rotation of the axes in the case of resonance offset. To the best of our knowledge, this is the first work that introduces a canonical version of the Bloch Equations and furthermore, that applies the concepts of classical stability analysis and caustics to the investigation of dynamics on the Bloch Sphere.

\subsection{Dynamics of two-level system}
The evolution of a quantum two-level system is given by the Liouville-von-Neumann Equation (LvN) ($\hbar=1$ is considered throughout work):
\begin{equation}
\label{LvN}
\frac{d\hat\rho}{dt}=-i[\hat H,\hat\rho],
\end{equation}
where $\hat\rho$ is a $2\times2$ density matrix:
\begin{equation}
\label{density}
\hat{\rho}=|b\rangle\langle a|=\begin{pmatrix}\rho_{bb} && \rho_{ba} \\ \rho_{ab} && \rho _{aa}\end{pmatrix},
\end{equation}
and $\hat H$ is the quantum Hamiltonian. The following two-level Hamiltonian is employed:
\begin{equation}
\label{quantum_Hamiltonian}
\hat H=\frac{1}{2}\Omega_{1}(t)\hat \sigma_{x}+\frac{1}{2}\omega_{0}\hat\sigma_{z}=\frac{1}{2}\begin{pmatrix}\omega_{0} && \Omega_{1}(t) \\ \Omega_{1}(t) && -\omega_{0}\end{pmatrix},
\end{equation}
where $\omega_{0}=\gamma B_{0}$ is the Larmor frequency and $\Omega_{1}(t)=\gamma B_{1}(t)$ is an RF field polarized along the $x$-axis, with $\gamma$ the gyromagnetic ratio of the nucleus; the field's carrier frequency is denoted $\omega$. In the optical nomenclature, $\omega_{0}=\epsilon_{b}-\epsilon_{a}$ is the level spacing between states $|a\rangle,|b\rangle$ and  $\Omega_{1}(t)$ plays the role of the time-dependent coupling $V_{ab}(t)=-\mu_{ab}\mathcal{E}(t)$, arising from the dipole interaction.\cite{Tannor2007IntroductionPerspective}

The Lie algebra $\mathfrak{su}(2)$, associated with the Lie group $\text{SU}(2)$, the symmetry group that corresponds to spin-1/2 particles, is isomorphic to the Lie algebra $\mathfrak{so}(3)$, associated with the Lie group of $\text{SO}(3)$ group, the rotation group in 3D.\citep{Kuprov2023Spin} In the latter representation, the matrix $\hat\rho$ is replaced by the so-called Bloch vector $\textbf{r}$:
\begin{equation}
\label{r-transformation}
\begin{cases}
x=\rho_{ba}+\rho_{ab} \\
y=i(\rho_{ba}-\rho_{ab}) \\
z=\rho_{bb}-\rho_{aa}
\end{cases}\hspace{0.5cm}.
\end{equation}
In a similar manner, the Hamiltonian is replaced by the $3$-vector $\boldsymbol{\Omega}$:
\begin{equation}
\label{Omega-transformation}
\begin{cases}
\Omega_{x}=\Omega_{1}^{*}+\Omega_{1} \\
\Omega_{y}=i(\Omega_{1}^{*}-\Omega_{1}) \\
\Omega_{z}=\omega_{0}
\end{cases}\hspace{0.5cm},
\end{equation}
and the commutator is replaced by a cross-product.\citep{Levitt1982SymmetricalOffset} Altogether, we obtain the so-called Bloch Equations (or the Optical Bloch Equations):
\begin{equation}
\label{Cartesian_Bloch_equations}
\dot{\textbf{r}}=\boldsymbol{\Omega}\times \textbf{r},
\end{equation}
which are the EOMs for the angular momentum components of a rotating rigid body. In this sense, one can understand $\dot{\textbf{r}}$ in eq. \ref{Cartesian_Bloch_equations} as analogous to a classical torque. The constraint $|\textbf{r}|=\sqrt{x^2+y^2+z^2}\leq 1$ in $\text{SO(3)}$ is equivalent to the condition $\text{tr}\hat\rho^{2}\leq1$ in $\text{SU(2)}$; pure states are characterized by $|\textbf{r}|=1$ in $\text{SO(3)}$ and $\text{tr}\hat\rho^{2}=\text{tr}\hat\rho=1$ in $\text{SU(2)}$. Therefore, any time-evolving pure state can be visualized as a trajectory on the so-called Bloch Sphere of radius $1$. In this representation, first proposed in its optical context by Feynman, Vernon, and Hellwarth,\citep{Feynman1957GeometricalProblems} the south pole corresponds to the state $|a\rangle$ and the north pole corresponds to the state $|b\rangle$. The spin states "up" ($|\uparrow\rangle$) and "down" ($|\downarrow\rangle$) play the role of the electronic ground state and the excited state. Any other point on the sphere represents a superposition state, and points inside the sphere, which are excluded in this work, represent mixed states. In terms of quantum computing, the entire Bloch Sphere is called a qubit. Without loss of generality, the initial state in this work is taken to be at the north pole of the Bloch Sphere, and the desired final state is at the south pole.

Eqs. \ref{Cartesian_Bloch_equations} can be also written as:
\begin{equation}
\label{Matrix-Bloch-equations}
\dot{\textbf{r}}=\boldsymbol{\underline{\underline{\Omega}}}\textbf{r},
\end{equation}
where now $\underline{\underline{\boldsymbol{\Omega}}}$ is a skew-symmetric matrix that represents the cross product of $\textbf{r}$ with the Cartesian components of $\boldsymbol{\Omega}$. Since $\underline{\underline{\boldsymbol{\Omega}}}$ is generally time-dependent, the formal solution of eqs. \ref{Matrix-Bloch-equations} is given by:
\begin{equation}
\label{formal_solution_Bloch}
\textbf{r}(t)=\hat{\mathbb{T}}\exp\left({\int_{t_{\text{i}}}^{t}\underline{\underline{\boldsymbol{\Omega}}}(t')dt'}\right)\textbf{r}_{\text{i}},
\end{equation}
where $\hat{\mathbb{T}}$ is the Dyson time-ordering operator and $\textbf{r}_{\text{i}}\equiv\textbf{r}(t_{\text{i}})$ represents the initial quantum state. If the fields entering into $\underline{\underline{\boldsymbol{\Omega}}}$ are monochromatic with constant amplitude, or at least piecewise-constant, then one can find a unitary transformation that will make eqs. \ref{Matrix-Bloch-equations} autonomous, i.e., that $\underline{\underline{\boldsymbol{\Omega}}}$ will be time-independent. Because $\underline{\underline{\boldsymbol{\Omega}}}$ includes multiple frequencies associated with the internal and external components of the Hamiltonian, the above transformation is possible only if the excitation is near-resonant, i.e.,
\begin{equation}
\label{RWA}
|\omega-\omega_{0}|\ll\omega+\omega_{0},
\end{equation}
such that emergent terms like $e^{+i(\omega+\omega_{0})t}$ will be negligible due to fast oscillations; this is the so-called Rotating Wave Approximation (RWA).\citep{Tannor2007IntroductionPerspective} The resonance offset $\omega-\omega_{0}$ is denoted $\Delta$, and the norm of $\boldsymbol{\Omega}$ is called the Rabi Frequency:
\begin{equation}
\label{Rabi}
\Omega=\sqrt{\Omega_{1}^{2}+\Delta^{2}},
\end{equation}
where ${\Omega}_{1}$ is the magnitude of the external field. The original time-dependent frame is denoted the Lab Frame, while the time-independent frame is denoted the Rotating Frame, since it rotates at the Rabi Frequency. In the Rotating Frame, the vector $\boldsymbol{\Omega}$ plays the role of a fixed rotation axis around which the Bloch vector precesses, and the Dyson time-ordering operator is replaced by an identity operator. The transformation to the Rotating Frame conserves the structure of the Bloch Equations, such that they are fulfilled in each segment separately, with the final state of one segment serving as the initial condition for the next segment.

\subsection{Composite pulses}
When CPs in the Rotating Frame are visualized as a function of time, since the carrier frequency is removed, they appear as a sequence of rectangles.\citep{Levitt1986CompositePulses} Each rectangular segment, enumerated $k$, in general possesses a characteristic amplitude, duration $\tau_{k}$, and phase $\varphi_{k}$; the latter determines the azimuthal angle of the rotation axis. Experimentally, it is desired to diminish the overall duration of free precession, such that the general notation is $\tau_{1}(\hat{n}_{1})\tau_{2}(\hat{n}_{2})...\tau_{N}(\hat{n}_{N})$ represents a series of back-to-back rotations, where $\hat{n}_{k}$ are the fixed rotation axes, given by:
\begin{equation}
\label{rotation_axis}
\hat{n}_{k}=n_{x}\hat x+n_{y}\hat y+n_{z}\hat z=\sin\vartheta_{k}\cos\varphi_{k}\hat{x}+\sin\vartheta_{k}\sin\varphi_{k}\hat{y}+\cos\vartheta_{k}\hat{z},
\end{equation}
where $\vartheta_{k}$ are the polar angles. Also, $\tau_{k}=t_{k}/T$ where $t_{k}$ is the clock time for the $k$-th pulse and $T=2\pi/\Omega_{1}^{(0)}$. As a consequence, we obtain an elegant formula for the overall effect of a CP on an initial state, which also provides the formal solution for the Bloch Equations in the Rotating Frame:
\begin{equation}
\label{rotation-solution}
\textbf{r}_{\text{f}}=\prod_{k=1}^{N}\mathcal{R}_{\hat{n}_{k}}(\beta_{k})\cdot \textbf{r}_{\text{i}},
\end{equation}
where $\textbf{r}_{\text{f}}\equiv\textbf{r}(t_{\text{f}})$ is the final state, and each $\mathcal{R}_{\hat{n}_{k}}$ is a general rotation matrix\citep{Kuprov2023Spin}
\begin{equation}
\label{general-rotation}
\mathcal{R}_{\hat{n}_{k}}(\beta_{k})=\exp\left[-\beta_{k}\begin{pmatrix}0 &&-n_{z} &&n_{y}\\n_{z}&&0 && -n_{x} \\-n_{y} && n_{x} &&0\end{pmatrix}\right].
\end{equation}
Note that the order of the $\mathcal{R}$ operators in the product is significant, i.e. $\mathcal{R}_{\hat{n}_{2}}$ operates after $\mathcal{R}_{\hat{n}_{1}}$, etc. Here, $\beta_{k}$ is the so-called flip angle; for ideal ensembles, $\beta_{k}=\tau_{k}$, i.e., the flip angles are precisely equal to the segment durations. We will discuss the deviations from this statement later in this section. Note that eq. \ref{general-rotation} presents a matrix exponential, which can be approximated by Rodrigues' Rotation Formula; however, in numerical implementation this is unneeded, and the matrix exponential is calculated with a much better precision. All in all, the time-evolution is now described by a series of rotations, and this is all allowed since $\underline{\underline{\boldsymbol{\Omega}}}$ is piecewise-constant.

The CP conceived by Levitt and Freeman is a symmetric sequence comprised of three rectangular segments.\citep{Levitt1979NMRPulse} In Levitt's nomenclature, this sequence is denoted $90^{\circ}(x)180^{\circ}(y)90^{\circ}(x)$, which we interpret as three sequential rotations: (1) around the $x$-axis, then (2) around the $y$-axis and eventually (3) around the $x$-axis again; this notation corresponds to the choice $\vartheta_{k}=\pi/2,\forall k$  in eq. \ref{rotation_axis}. To prevent any confusion, we rewrite this sequence in the notation of eq. \ref{rotation-solution}:
\begin{equation}
\label{Levitt-pulse}
  \textbf{r}_{\text{f}}=\mathcal{R}_{x}(T/4)R_{y}(T/2)R_{x}(T/4)\cdot\textbf{r}_{\text{i}}.
\end{equation}
Levitt found that this pulse sequence can compensate for two imperfections of the Hamiltonian in eq. \ref{quantum_Hamiltonian}: (1) field inhomogeneity and (2) resonance offset. In case (1), each member of the ensemble is driven by a slightly different external field, i.e., if the magnitude of the nominal external field is $\Omega_{1}^{(0)}=1$, then the magnitude of the field felt by the $j$-th member is $\Omega_{1}^{(j)}$. In case (2), each member of the ensemble has a slightly different offset $\Delta^{(j)}=\omega-\omega_{0}^{(j)}$. Following Levitt's work, various extensions have been suggested, many of them consisting of more than just three segments; Levitt himself suggested a five-segment sequence that outperformed his original sequence (eq. \ref{Levitt-pulse}).\citep{Levitt1982SymmetricalOffsetb}

\begin{figure}[ht!]
    \centering
    \includegraphics[width=1\linewidth]{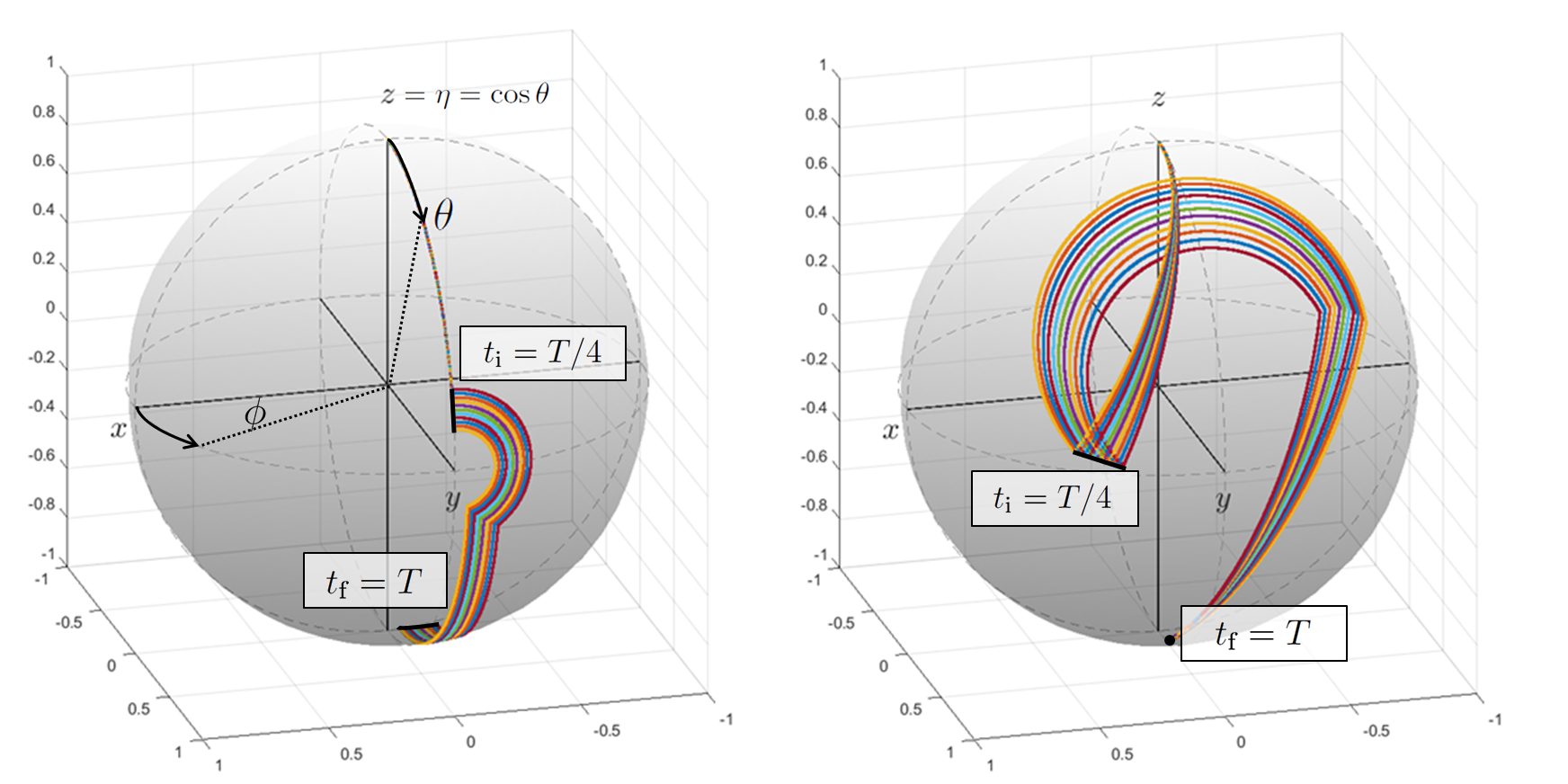}
    \caption{A set of trajectories on the Bloch Sphere, representing the time-evolution of an ensemble under the Levitt pulse sequence $90^{\circ}(x)180^{\circ}(y)90^{\circ}(x)$ with two characteristic imperfections: field inhomogeneity, ranging from $0.8$ to $0.9$ with respect to the nominal field magnitude (left) and resonance offsets ranging from $0.4$ to $0.6$ (right). The solid black lines at $t_{\text{i}}=/T/4$ and $t_{\text{f}}=T$ mark the initial and the final manifold discussed along this work.}
    \label{fig:Levitt_recovered}
\end{figure}

These two imperfections lead to variations in the overall effect of the pulse on each member of the ensemble; the effect is two-fold - on the effective rotation axis and the effective flip angle:
\begin{equation}
\label{theta-beta}
\vartheta^{(j)}=\tan^{-1}\left(\frac{\mathcal E_{0}^{(j)}}{\Delta^{(j)}}\right), \hspace{0.5cm} \beta_{k}^{(j)}=\tau_{k}\left({\frac{\Omega^{(j)}}{\Omega_{1}^{(0)}}}\right).
\end{equation}
Note that $\vartheta^{(j)}$ does not contain the subscript $k$, since all segments have the same magnitude (in absolute value) for a certain member $j$ of the ensemble. If $\Delta^{(j)}=0$, the axis of rotation will point along the equator. Otherwise, it will incline or decline a bit in the $z$ direction. In this work, we will treat imperfections (1) and (2) separately; this isolation of variables will put our findings in a clearer context. Fig. \ref{fig:Levitt_recovered} shows the effect of these two imperfections. 

Since this Rotating Frame formulation (eq. \ref{rotation-solution}) does not contain time as an explicit continuous parameter, one may incorporate it artificially into the numerical simulation of eq. \ref{Levitt-pulse}, in order to obtain a trajectory evolving on the Bloch Sphere; otherwise, one obtains only the endpoints of each segment. Due to unitarity, the real time-evolution in the Lab Frame can be recovered by artificially slicing the dynamics in the Rotating Frame into "moments"; this slicing requires cautious bookkeeping such that one can map the artificial moments from the Rotating Frame to the real-time in the Lab Frame.

Levitt found that his pulse sequence can compensate for both imperfections, as seen by the endpoints of the trajectories, which are strikingly close to the antipode of the starting point. We note that these cases are rather different: in case (1), the endpoints are spread along the azimuth $\phi$ and focused in the polar angle $\theta$, while in case (2) the focusing is apparent in both $\theta$ and $\phi$. We will return to this difference later in this work.

In one of his later works,\citep{Levitt1982SymmetricalOffset,Levitt1982SymmetricalOffsetb} Levitt explained the successful performance of his three-segment sequence by considering the imperfections as perturbations with respect to the final state of a nominal representative of the ensemble, which experiences $\Omega_{1}^{(0)}$ as-is and its internal frequency $\omega_{0}$ is on-resonance with $\omega$. In this perturbative argument, the Bloch vector is expanded as a Taylor series:
\begin{equation}
\label{Levitt-expansion}
\textbf{r}_{\text{f}}=\textbf{r}_{\text{f}}^{(0)}+\left(\frac{\partial \textbf{r}_{\text{f}}}{\partial w} \right)_{w=w^{(0)}}\Delta w+\mathcal{O}(\Delta w^{2}),
\end{equation}
where $w$ is one of the Cartesian components of $\boldsymbol\Omega$; for a real-value external field in the Rotating Frame, $w=\Omega_{1}^{(j)}$ or $\Delta^{(j)}$. $\Delta w$\footnote{note that this notation shall be read as one piece.} represents a small increment in one of these components, i.e., the imperfection. If the quantity $\left(\frac{\partial \textbf{r}_{\text{f}}}{\partial w} \right)$ is minimized, then the resulting error due to imperfection is minimized as well, and the overall compensation effect due to the robust pulse is considered good. In this work, we connect this argument with classical stability analysis.

Throughout this work, we consider: $\Omega_{1}^{(0)}=\omega=1$, $\Omega_{1}^{(j)}\in[0.8,0.9]\Omega_{1}^{(0)}$ and $\Delta^{(j)}\in[0.4,0.6]$ with $j=1,2,...,101$ and $j=1,2,...,201$ respectively; $10^{5}$ time steps were taken per each trajectory; as said, all trajectories start from the north pole. Convergence of trajectories is verified with respect to the time step $dt$, and stability calculations are converged with respect to the numerical steps $d{\Omega}_{1}$ and $d\Delta$. Numerical differentiations are done with finite differences.

\section{Derivations and methods: a new formulation for Bloch Equations}

\subsection{Non-canonical coordinates}
Since we are using $\text{SO(3)}$ as a surrogate for $\text{SU(2)}$, it is well to revisit the physical interpretation of the Bloch vector $\textbf{r}$. In classical mechanics, the functions $f_{\mu}(\textbf{q},\textbf{p},t)$ and $g_{\nu}(\textbf{q},\textbf{p},t)$ define a canonical pair if they provide the following Poisson bracket relation ($\textbf{q}$ and $\textbf{p}$ have the usual meaning of position and momentum, respectively:
\begin{equation}
\label{general-Possion}
\{f_{\mu},g_{\nu}\}\equiv\sum_{i=1}^{N}\left(\frac{\partial f_{\mu}}{\partial q_{i}}\frac{\partial g_{\nu}}{\partial p_{i}}-\frac{\partial f_{\mu}}{\partial p_{i}}\frac{\partial g_{\nu}}{\partial q_{i}}\right)=\delta_{\mu\nu},
\end{equation}
where $\delta_{\nu\mu}$ is the Kronecker delta; in particular, if $\{q_{\nu},p_{\nu}\}=1$, then the position coordinate and its conjugate momentum are canonical.

Naïvely, one might have considered $\textbf{r}$ as a classical position vector, trying to treat it in the view of eq. \ref{general-Possion}; however, note that the components of $\textbf{r}$ satisfy the following Poisson bracket relation ($\mu,\nu,\lambda=1,2,3$):
\begin{equation}
\label{cartesian-Poisson}
\{r_{\mu},r_{\nu}\}=-\epsilon_{\mu\nu\lambda}r_{\lambda}\neq\delta_{\mu\nu},
\end{equation}
where $\epsilon_{\mu\nu\lambda}$ is the Levi-Civita symbol. The components of $\textbf{r}$ satisfy a special algebraic structure, called the Lie-Poisson structure;\citep{Marsden1999IntroductionSymmetry} together with the constraint of pure states to live on the Bloch Sphere, the components of $\textbf{r}$ are not independent and therefore are non-canonical. We noted above that EOMs for $\textbf{r}$ are identical to those of the angular momentum components of a rotating rigid body, and indeed, eq. \ref{cartesian-Poisson} is satisfied by the components of orbital angular momentum. In the case of orbital angular momentum, it can be explicitly verified that $\{\ell_{\mu},\ell_{\nu}\}=\epsilon_{\mu\nu\lambda}\ell_{\lambda}\neq\delta_{\mu\nu}$ because each of these components are defined in terms of $\textbf{q}$ and $\textbf{p}$. In the case of the Bloch vector $\textbf{r}$, eq. \ref{general-Possion} cannot be evaluated explicitly, because $\textbf{r}$ is not a function of $\textbf{q}$ and $\textbf{p}$, but by analogy with the angular momentum vector, again $\{r_{\mu},r_{\nu}\}=-\epsilon_{\mu\nu\lambda}\ell_{\lambda}\neq\delta_{\mu\nu}$.

\subsection{Defining the Hamiltonian for non-canonical coordinates
}At this point, we aim to define a classical Hamiltonian, which corresponds to rotational energy. In classical mechanics, rotational energy is given by the product of angular frequency $\boldsymbol{\Omega}$ and angular momentum $\textbf{r}$:
\begin{equation}
\label{cartesian-classical-Hamiltonian}
H(x,y,z)=\boldsymbol{\Omega}\cdot \textbf{r} =\Omega_{x}x+\Omega_{y}y+\Omega_{z}z.
\end{equation}
This Hamiltonian resembles the quantum Hamiltonian in eq. \ref{quantum_Hamiltonian}, where now the components of $\textbf{r}$ play the role of Pauli matrices; note that for real-valued electric fields, i.e., $\boldsymbol{\Omega}=(-\Omega_{1},0,\omega_{0})$, both the quantum and the classical Hamiltonians have the structure of $x$-term plus $z$-term. In this non-canonical representation, the components of $\textbf{r}$ cover the entire phase space; there are no variables conjugate to the components of $\textbf{r}$. Thus, the phase space consists of only 3 degrees of freedom. This Hamiltonian is sometimes called the Zeeman Hamiltonian, due to its connection to the Zeeman effect, where a magnetic field interacts with the total angular momentum of a quantum system.\citep{Kuprov2023Spin}

The EOMs for $\textbf{r}$ have the following form, which is a general Lie-Poisson structure:
\begin{equation}
\label{Lie-Poisson}
\dot{\textbf{r}}={}\omega(\textbf{r})\cdot\boldsymbol{\nabla} H(\textbf{r})+\frac{\partial \textbf{r}}{\partial t}.
\end{equation}
The matrix $\omega(\textbf{r})$ contains the relations between the components of the $\textbf{r}$; in this case, it will be the following skew-symmetric matrix:
\begin{equation}
\label{skew-symmetric}
\omega(\textbf{r})=
\begin{pmatrix}
0 && -z && y\\
z&& 0&&-x \\
-y && x && 0
\end{pmatrix}=-\textbf{r}\times \cdot\hspace{0.5cm}.
\end{equation}
Since
\begin{equation}
\label{grad-and-time-derivative}
\boldsymbol{\nabla} H=\frac{\partial H}{\partial \textbf{r}}=\boldsymbol{\Omega},\hspace{0.5cm}\frac{\partial \textbf{r}}{\partial t}=0,
\end{equation}
we obtain the following EOMs:
\begin{equation}
\label{Bloch-auxiliary}
\dot{\textbf{r}}=-\textbf{r}\times\boldsymbol{\Omega}=\boldsymbol{\Omega}\times \textbf{r},
\end{equation}
which are the Bloch Equations, now understood as Hamilton's Equations for the Zeeman Hamiltonian. In this context, we note that a unitary transformation from the Lab Frame to the Rotating Frame can be considered a canonical transformation since it preserves the structure of Hamilton's Equations. With this understanding, we see why the Bloch Equations represent a unique situation: we are able to solve a quantum problem by  mapping it to a classical problem; the classical solutions may be mapped to give the exact quantum solutions with no approximations.

Note that $\omega(\textbf{r})$ is a generalization of the two-form $dq\wedge dp$; for Hamiltonians in canonical form, $\omega(\textbf{r})$ is the usual symplectic matrix $\mathbb{J}=\begin{pmatrix}0 && \mathbb{I}_{n} \\ -\mathbb{I}_{n} && 0\end{pmatrix}$, which is independent of $\textbf{r}$.\citep{Ross2020Poisson12} 

\subsection{Canonical coordinates}
Now that we have found the classical Hamiltonian in terms of Cartesian coordinates, we will change variables to spherical coordinates, a natural choice for a physical system that lives on a sphere. Given the conventional transformation from Cartesian to spherical coordinates:\citep{Slichter1990PrinciplesResonance}
\begin{equation}
\label{cart-to-sph}
\begin{cases}
x=R\sin\theta\cos\phi \\
y=R\sin\theta\sin\phi \\
z=R\cos\theta
\end{cases},\hspace{0.5cm}
\end{equation}
where $R=1$ in the case of pure states, the Hamiltonian in spherical coordinates is
\begin{equation}
\label{spherical-classical-Hamiltonian}
H(\theta,\phi)=\Omega_{x}\sin\theta\cos\phi+\Omega_{y}\sin\theta\sin\phi+\Omega_{z}\cos\theta.
\end{equation}
The Cartesian components of $\boldsymbol{\Omega}$ are left as-is intentionally; transforming them to a spherical representation is unnecessary since only the coordinates are of interest to us. Now, we make the following additional change of variables:
\begin{equation}
\label{eta-theta}
\eta\equiv z=\cos\theta\Rightarrow\sqrt{1-\eta^2}=\sin\theta,
\end{equation}
such that
\begin{equation}
\label{canonical-classical-Hamiltonian}
H(\phi,\eta)=\Omega_{x}\sqrt{1-\eta^2}\cos\phi+\Omega_{y}\sqrt{1-\eta^{2}}\sin\phi+\Omega_{z}\eta;
\end{equation}
this reduces the phase space to 2 degrees of freedom instead of the original 3. We may identify the structure of an internal Hamiltonian $H_{0}=\Omega_{z}\eta$  and an external field $H_{\text{ext}}=\Omega_{x}\sqrt{1-\eta^2}\cos\phi+\Omega_{y}\sqrt{1-\eta^{2}}\sin\phi$, analogous to the structure of the original quantum Hamiltonian, eq. \ref{quantum_Hamiltonian}. This classical Hamiltonian is integrable if $\boldsymbol\Omega$ is time-independent, and the energy is conserved according to Noether's Theorem; for CPs, with their piecewise time-dependence, the energy will be piecewise-conserved. If $\boldsymbol\Omega$ has general time-dependence, then the EOMs are non-autonomous ("1.5D"), and the dynamics might be chaotic under certain circumstances. Previous works showed that this is indeed the case, e.g. if the external field is two-color with incommensurate frequencies chaos can ensue.\citep{Eidson1986QuantumField,Pomeau1986ChaoticPerturbation}

The variables $\phi$ and $\eta$ are in fact canonical. To show this, we calculate their Poisson brackets:
\begin{equation}
\label{phi-eta-Poisson-1}
\{\phi,\eta\}=\frac{\partial \phi}{\partial x}\{x,\eta\}+\frac{\partial \phi}{\partial y}\{y,\eta\}=\frac{\partial\phi}{\partial x}y-\frac{\partial\phi}{\partial y}x=1,
\end{equation}
where we have used eqs. \ref{Lie-Poisson}, \ref{cart-to-sph}, \ref{eta-theta}. We have chosen $\phi$ as the position and $\eta$ as its conjugate momentum; we will show later why this choice is natural. In terms of the new canonical coordinates, Hamilton's Equations take the form:
\begin{equation}
\label{canonical-Bloch}
\begin{cases}
\dot{\phi}=\frac{\partial H}{\partial \eta}=-\Omega_{x}\frac{\eta} {\sqrt{1-\eta^{2}}}\cos\phi-\Omega_{y}\frac{\eta} {\sqrt{1-\eta^{2}}}\sin\phi+\Omega_{z}\\
\dot{\eta}=-\frac{\partial H}{\partial \phi}=\Omega_{x}\sqrt{1-\eta^{2}}\sin\phi-\Omega_{y}\sqrt{1-\eta^{2}}\cos\phi
\end{cases}\hspace{0.5cm},
\end{equation}
whose solutions are equivalent to the solutions of the Cartesian Bloch Equations. However, the compact representation of two-level dynamics in eqs. \ref{canonical-Bloch} suffers from a singularity: at the poles ($\eta=\pm1$), which are extremely meaningful in this work, $\phi$ is not well-defined; also, $\dot{\phi}$ diverges and $\dot{\eta}$ vanishes there, requiring careful numerical attention around these points.

\begin{figure}[ht!]
    \centering
    \includegraphics[width=1\linewidth]{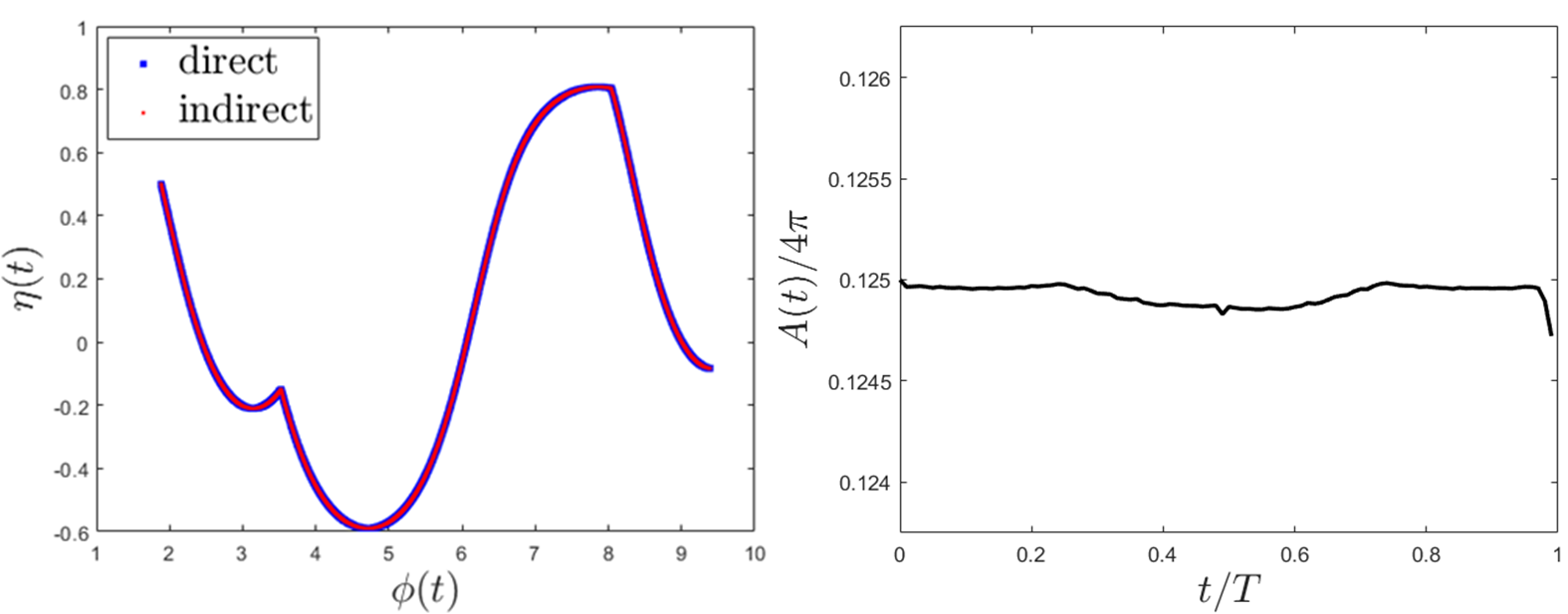}
    \caption{Left: phase space representation of a trajectory starting from $(\phi_{0},\eta_{0})=(0.6\pi,0.5)$, evolving under Levitt's pulse sequence (note that the unwrapped time-series $\phi(t)$ is plotted). The solution obtained by direct integration of the canonical Bloch Equations \ref{canonical-Bloch}, and via an indirect calculation based on an integration of the Cartesian Bloch Equations are in excellent agreement; such agreement is obtained for any initial conditions, except $\eta=\pm1$. Right: the time-evolution of the populated area (classical density) in phase space, starting from an initially rectangular population (), showing that Liouville's theorem is satisfied within numerical precision.}
    \label{fig:agreement_and_Liouville}
\end{figure}

A bypass for the singularity issue is to obtain the trajectories with the Cartesian Bloch Equations (eqs. \ref{Cartesian_Bloch_equations}), then to calculate $\phi(t)=\tan^{-1}\left(\frac{y(t)}{x(t)}\right)$, and trivially, $\eta(t)=z(t)$. It can be shown that both the Cartesian components of $\textbf{r}$ and the new canonical coordinates satisfy Liouville's theorem. Liouville's theorem expresses itself in terms of the conservation of area $A$ on the surface of the Bloch Sphere. Figure \ref{fig:agreement_and_Liouville} shows the time evolution of the populated area in phase space in the case of canonical coordinates; the small deviations from $A(t)=\pi/2$ are due to numerical noise.

\subsection{The average stability elements}
The canonical position and momentum enable us to discuss the effect of the CPs in terms of stability. Since Hamiltonian flow is area-preserving, both in the original Cartesian coordinates and in the canonical coordinates, we choose to work with canonical coordinates as a matter of convenience.

In classical mechanics, the stability matrix (or monodromy matrix) describes the stability of the EOMs' solutions under small changes in the initial variables. For instance, in a general 1D problem, this matrix, denoted $\mathcal{M}$, is\citep{Tannor2007IntroductionPerspective}
\begin{equation}
\label{stability-prototype}
\mathcal{M}=\begin{pmatrix}
\frac {\partial p_{\text{f}}}{\partial p_{\text{i}}} & \frac {\partial p_{\text{f}}}{\partial q_{\text{i}}} \\
\frac{\partial q_{\text{f}}}{\partial p_{\text{i}}}  &  \frac {\partial q_{\text{f}}}{\partial q_{\text{i}}}
\end{pmatrix},
\end{equation}
which expresses the changes in the "final" variables (denoted with subscript $\text{f}$) resulting from small changes in the "initial" variables (denoted with subscript $\text{i}$). The changes are intimately associated with the so-called tangent space, which is orthogonal to the usual view of Hamiltonian time-evolution in terms of trajectories.
The eigenvalues of the stability matrix are the precursors of the Lyapunov Exponents; the latter indicate the emergence of chaos versus long-time regularity. Moreover, the stability matrix elements themselves are indicative of full or partial refocusing in different directions of the phase space, and therefore are of interest to us.

We now turn to the description of the ensemble of trajectories on the Bloch Sphere using the terminology of classical mechanics. Often, one is interested in an ensemble that can be described by a manifold embedded in the full phase space, the so-called Lagrangian Manifold, a global object that drifts in phase space under the Hamiltonian flow. One of the striking features of a swarm of trajectories is the appearance of so-called caustic. This is a geometrical property of an evolving manifold, which appears when many (or all) initial conditions lead to the same final condition. In the case of a two-dimensional phase space, this is the condition that $\partial x_{\text{f}}/\partial p_{\text{i}}=0$ or in our case that $\partial\phi_{\text{f}}/\partial\eta_{\text{i}}=0$.\citep{Littlejohn1992TheGeometry}

The canonical framework of eq. \ref{canonical-Bloch} allows an immediate implementation of eq. \ref{stability-prototype}, where $q=\phi$ and $p=\eta$. Formally, we also need to include $R$, to obtain the following stability matrix:
\begin{equation}
\label{stability-spherical}
\mathcal{M}_{\text{s}}=\begin{pmatrix}\partial R_{\text{f}}/\partial R_{\text{i}} && \partial R_{\text{f}}/\partial\phi_{\text{i}} && \partial R_{\text{f}}/\partial\eta_{\text{i}} \\ \partial \phi_{\text{f}}/\partial R_{\text{i}} && \partial \phi_{\text{f}}/\partial \phi_{\text{i}}  && \partial \phi_{\text{f}}/\partial\eta_{\text{i}} \\ \partial\eta_{\text{f}}/\partial R_{\text{i}} && \partial\eta_{\text{f}}/\partial\phi_{\text{i}} && \partial\eta_{\text{f}}/\partial\eta_{\text{i}}\end{pmatrix}=\begin{pmatrix}1 && 0 && 0 \\ 0 && \partial \phi_{\text{f}}/\partial \phi_{\text{i}}  && \partial \phi_{\text{f}}/\partial\eta_{\text{i}} \\ 0 && \partial\eta_{\text{f}}/\partial\phi_{\text{i}} && \partial\eta_{\text{f}}/\partial\eta_{\text{i}}\end{pmatrix},
\end{equation}
where we have set $R=1$; the subscript "s" stands for "spherical". Clearly, there are only four stability elements, such that the effective stability matrix is $2\times2$. To find the relationship between the spherical stability matrix $\mathcal{M}_{\text{s}}=\left(\frac{\partial \textbf{r}_{\text{f}}}{\partial \textbf{r}_{\text{i}}}\right)_{\text{s}}$ and the Cartesian stability matrix $\mathcal{M}_{\text{c}}\equiv\left(\frac{\partial \textbf{r}_{\text{f}}}{\partial \textbf{r}_{\text{i}}}\right)_{\text{c}}$ we use the Jacobian $\mathcal{J}$ that transforms between the coordinate systems:
\begin{equation}
\label{chain-rule}
\mathcal{M}_{\text{c}}\equiv\left(\frac{\partial \textbf{r}_{\text{f}}}{\partial \textbf{r}_{\text{i}}}\right)_{\text{c}}=\left(\frac{\partial (x_{\text{f}},y_{\text{f}},z_{\text{f}})}{\partial (R_{\text{f}},\phi_{\text{f}},\eta_{\text{f}})}\right)\left(\frac{\partial \textbf{r}_{\text{f}}}{\partial \textbf{r}_{\text{i}}}\right)_{\text{s}}\left(\frac{\partial (R_{\text{i}},\phi_{\text{i}},\eta_{\text{i}})}{\partial (x_{\text{i}},y_{\text{i}},z_{\text{i}})}\right)\equiv\mathcal{J}_{\text{f}}\mathcal{M}_{\text{s}}\mathcal{J}_{\text{i}}^{-1},
\end{equation}
i.e.
\begin{equation}
\label{Jacobian-f}
\mathcal{J}_{\text{f}}=\begin{pmatrix}\partial x_{\text{f}}/\partial R_{\text{f}} &&  \partial x_{\text{f}}/\partial \phi_{\text{f}}  && \partial x_{\text{f}}/\partial \eta_{\text{f}}  \\ \partial y_{\text{f}}/\partial R_{\text{f}} &&  \partial y_{\text{f}}/\partial \phi_{\text{f}}  && \partial y_{\text{f}}/\partial \eta_{\text{f}} \\ \partial z_{\text{f}}/\partial R_{\text{f}} &&  \partial z_{\text{f}}/\partial \phi_{\text{f}}  && \partial z_{\text{f}}/\partial \eta_{\text{f}}\end{pmatrix}=\begin{pmatrix}\cos\phi_{\text{f}}\sqrt{1-\eta_{\text{f}}^2} &&  -\sin\phi_{\text{f}}\sqrt{1-\eta_{\text{f}}^2}  && 0  \\ \sin\phi_{\text{f}}\sqrt{1-\eta_{\text{f}}^2} &&  \cos\phi_{\text{f}}\sqrt{1-\eta_{\text{f}}^2}  &&  0 \\ \eta_{\text{f}} &&  0  && 1\end{pmatrix},
\end{equation}
and
\begin{equation}
\label{Jacobian-i}
\mathcal{J}_{\text{i}}^{-1}=\begin{pmatrix}\partial R_{\text{i}}/\partial x_{\text{i}} &&  \partial R_{\text{i}}/\partial y_{\text{i}}  && \partial R_{\text{i}}/\partial z_{\text{i}}  \\ \partial \phi_{\text{i}}/\partial x_{\text{i}} &&  \partial \phi_{\text{i}}/\partial y_{\text{i}}  && \partial \phi_{\text{i}}/\partial z_{\text{i}} \\ \partial \eta_{\text{i}}/\partial x_{\text{i}} &&  \partial \eta_{\text{i}}/\partial y_{\text{i}}  && \partial \eta_{\text{i}}/\partial z_{\text{i}}\end{pmatrix}=\begin{pmatrix}\cos\phi_{\text{i}}\sqrt{1-\eta_{\text{i}}^2}&&  \sin\phi_{\text{i}}\sqrt{1-\eta_{\text{i}}^2} && -\eta_{\text{i}}  \\ -\sin\phi_{\text{i}} \sqrt{1-\eta_{\text{i}}^2}&& \cos\phi_{\text{i}}\sqrt{1-\eta_{\text{i}}^2}  &&  0 \\ 0 &&  0  && 1\end{pmatrix},
\end{equation}
where the upper-left blocks are rotation matrices.

Numerically, the calculation of the stability elements is done by considering small separations $\Delta\phi_{\text{i}},\Delta\eta_{\text{i}}$ and propagating a central swarm and the requisite satellite swarms: each trajectory evolves with its own Hamiltonian and the stability elements are obtained at each moment using finite-difference derivatives. This routine is based on the suggestion of Heller ($\zeta=\phi,\eta$):\citep{Heller1976ClassicalDynamicsb}
\begin{equation}
\label{Heller}
\frac{\partial\zeta_{\text{f}}}{\partial\zeta_{\text{i}}}\approx\frac{\Delta\zeta_{\text{f}}}{\Delta\zeta_{\text{i}}},
\end{equation}
where $\Delta\zeta_{\text{i}}$ and $\Delta\zeta_{\text{f}}$ are finite differences. Since we are not interested in characterizing chaotic motion, which in the case of simple CPs does not exist, we shall not carry out any renormalization scheme.

There are three challenges that come up here: (1) contrary to previous works in which Lagrangian Manifolds comprise many initial positions and/or momenta, we are interested in the case where all trajectories start from the same initial conditions at $t_{\text{i}}=0$; (2) contrary to the traditional stability analysis, where all trajectories evolve under the same Hamiltonian, here the Hamiltonian is characterized by slightly different parameters per each trajectory, making the analysis more challenging; (3) the straightforward implementation of the eqs. \ref{canonical-Bloch} precludes examining $\phi$ at the poles.

In order to overcome challenge (1) for the stability analysis, we take the initial time $t_{\text{i}}=T/4$, where the swarm features maximal spreading, and we consider $t_{\text{f}}\in[T/4,T]$. To distinguish $t_{\text{i}}$ from the moment when the trajectories actually start, we will denote the latter as $t_{0}=0$.

In order to overcome challenge (2), we define average stability matrix elements:
\begin{equation}
\label{average-stability}
\langle \mathcal{M}_{\text{s}}\rangle\equiv\left\langle\frac{\partial\zeta_{\text{f}}}{\partial{\zeta}_{\text{i}}}\right\rangle_{[w,w+\Delta w]}=\int_{w}^{w+\Delta w}\left({\frac{\partial\zeta_{\text{f}}}{\partial{\zeta}_{\text{i}}}}\right)dw'.
\end{equation}
The average stability elements are an excellent qualitative tool to characterize the evolution of the stability matrix elements of the entire ensemble.

We extend the ordinary framework of caustics, and conjecture that the average stability matrix elements shall indicate refocusing. If the refocusing happens close to the desired endpoint, this signifies that the ensemble has undergone a successful population inversion. In particular, if the histogram of the stability matrix element $\langle\partial\zeta_{\text{f}}/\partial\zeta_{\text{i}}\rangle$ collapses into a relatively narrow band, this indicates refocusing in the $\zeta$-direction. To measure the size of the band of certain stability element (even before averaging and the absolute value), we define the range parameter $h_{\zeta}$
\begin{equation}
\label{range-parameter}
h_{\zeta}(t_{\text{f}})=\max\left(\frac{\partial\zeta_{\text{f}}}{\partial\zeta_{\text{i}}}\right)-\min\left(\frac{\partial\zeta_{\text{f}}}{\partial\zeta_{\text{i}}}\right),
\end{equation}
which we will use as a supplemental measure of refocusing; consequently, we expect $h_{\zeta}$ to obtain its minimum at $t_{\text{f}}=T$. In other words, when we look for refocusing of a quantum ensemble, we are actually looking for a classical caustic.

In order to overcome challenge (3), in calculating $\partial\phi_{\text{f}}/\partial\phi_{\text{i}}$ we launch the central and satellite swarms of trajectories not from $\eta_{0}=1$ but from $\eta_{0}=1-\epsilon$ where $\epsilon$ is a parameter much smaller than the shift $\Delta\phi_{0}$.

We now recall Levitt's justification (eq. \ref{Levitt-expansion}) of his pulse sequence. Using the chain rule, we define the imperfection measure $\mathcal{W}$
\begin{equation}
\label{Levitt-chain-rule}
\mathcal{W}\equiv\left(\frac{\partial \textbf{r}_{\text{f}}}{\partial w}\right)=\left(\frac{\partial\textbf{r}_{\text{f}}}{\partial \textbf{r}_{\text{i}}}\right)\left(\frac{\partial \textbf{r}_{\text{i}}}{\partial w}\right)\approx\mathcal{M}_{\text{c}}\cdot\text{const}.= \mathcal{J}_{\text{f}}\mathcal{M}_{\text{s}}\mathcal{J}_{\text{i}}^{-1}\cdot\text{const}.,
\end{equation}
where we have used eq. \ref{chain-rule} and we have assumed that the dependence of $\textbf{r}_{\text{i}}$ on $w$ (where $w$ is either $\Omega_{1}^{(j)}$ or $\Delta^{(j)}$) is approximately linear. Eq. \ref{Levitt-chain-rule} provides a relation between $\mathcal{W}$ and the spherical stability matrix $\mathcal{M}_{\text{s}}$, within this assumption of linearity.

Recall that one needs to insert $w=w^{(0)}$ in eq. \ref{Levitt-expansion}; this insertion is equivalent to locally averaging over the elements of the spherical stability matrix. Inverting eq. \ref{Levitt-chain-rule} we obtain:
\begin{equation}
\label{overall-chain-rule-average}
\langle\mathcal{M}_{\text{s}}\rangle=\mathcal{J}_{\text{i}}\langle\mathcal{W}\rangle\mathcal{J}_{\text{f}}^{-1}\cdot\text{const.}
\end{equation}
Therefore, the caustics in the element $\mathcal{M}_{\text{s}}$, which are statistically observed by the collapse of the histogram, signal the minimization of Levitt's imperfection measure. 

The linearity of $\textbf{r}_{\text{i}}$ with respect to $w={\Omega}_{1}$ can be seen in fig. \ref{fig:Levitt_recovered} for $t_{\text{i}}=T/4$, particularly in the left subfigure. In the right subfigure, the linearity of $\textbf{r}_{\text{i}}$ with respect to $w=\Delta$ is less well satisfied. Since the connection between $\mathcal{M}_{\text{s}}$ and $\mathcal{W}$ might be specific to the ensemble of field inhomogeneity, we are cautious about using it to justify Levitt's pulse sequence for resonance offsets.

\subsection{The Lagrangian and canonical momentum}
Using the canonical coordinates and the Legendre transform, $\mathcal{L}=\eta\dot\phi-H$, we may define a Lagrangian. Using eqs. \ref{canonical-classical-Hamiltonian} and \ref{canonical-Bloch} we get:
\begin{equation}
\label{Lagrangian-1}
\mathcal{L}=\\-\frac{\Omega_{x}} {\sqrt{1-\eta^{2}}}\cos\phi-\frac{\Omega_{y}} {\sqrt{1-\eta^{2}}}\sin\phi.
\end{equation}
We may also express $\eta$ in terms of $\dot{\phi}$. Inverting the $\phi$-equation in eq. \ref{canonical-Bloch} we obtain:
\begin{equation}
\label{eta-phidot}
\begin{cases}
\eta=\frac{\Omega_{z}-\dot{\phi}}{\sqrt{(\Omega_{x}\cos\phi+\Omega_{y}\sin\phi)^2+(\Omega_{z}-\dot{\phi})^2}} \\
\frac{1}{\sqrt{1-\eta^{2}}}=\frac{\sqrt{(\Omega_{x}\cos\phi+\Omega_{y}\sin\phi)^2+(\Omega_{z}-\dot{\phi})^2}}{\Omega_{x}\cos\phi+\Omega_{y}\sin\phi}
\end{cases}\hspace{0.5cm},
\end{equation}
such that the Lagrangian has the following form:
\begin{equation}
\label{Lagrangian-final}
\mathcal{L}(\dot\phi,\phi)=-\sqrt{(\Omega_{z}-\dot{\phi})^2+(\Omega_{x}\cos\phi+\Omega_{y}\sin\phi)^2}.
\end{equation}
Identifying the $\dot\phi$-terms as those associated with the "kinetic energy" $K$ and the $\phi$-terms as those associated with the "potential energy" $V$: 
\begin{equation}
\label{kinetic-potential}
K(\dot\phi)=(\Omega_{z}-\dot\phi)^{2},\hspace{0.5cm}V(\phi)=-(\Omega_{x}\cos\phi+\Omega_{y}\sin\phi)^{2},
\end{equation}
the structure of $(K-V)^{1/2}$ is evident. Accordingly, the canonical momentum is
\begin{equation}
\label{canonical-momentum}
p_{\phi}=\frac{\partial\mathcal{L}}{\partial\dot\phi}=\pm\frac{\Omega_{z}-\dot\phi}{\sqrt{(\Omega_{z}-\dot{\phi})^2+(\Omega_{x}\cos\phi+\Omega_{y}\sin\phi)^2}}=\pm\eta,
\end{equation}
or, in other words, $|p_{\phi}|=|\eta|$. We see that $\eta$ is indeed a momentum variable as intuitively designated earlier. Numerical simulations show that the correct sign ("correct" here is compared to the direct integration of $\eta(t)=z(t)$ from the Cartesian Bloch Equations) alternates along the dynamics; the canonical momentum has two branches, and it hops between them throughout the time-evolution.

At this point, we differentiate the Lagrangian in order to obtain the Euler-Lagrange Equation. We assume that $\partial \mathcal{L}/\partial t\approx0$, which is valid if $\boldsymbol{\Omega}$ is piecewise-constant, as in CPs, except in a finite number of moments; if $\boldsymbol{\Omega}$ is strictly time-independent, this approximation is accurate and we obtain:
\begin{equation}
\label{EL-equation-reduced}
\ddot\phi=\frac{[(\Omega_x\cos\phi+\Omega_y\sin\phi)^2+(\Omega_{z}-\dot\phi)(\Omega_{z}-2\dot\phi)][(\Omega_y^2-\Omega_x^2)\sin2\phi+2\Omega_x\Omega_y\cos2\phi]}{2(\Omega_x\cos\phi+\Omega_y\sin\phi)^2}.
\end{equation}

Equation \ref{EL-equation-reduced} is easily checked to be equivalent to Hamilton's Equations (eq. \ref{canonical-Bloch}) under the assumption on the time-independence of the Lagrangian.

\section{Numerical results and analysis}

\subsection{The ensemble of field inhomogeneity }
Figure \ref{fig:Levitt_recovered} (left) shows the evolution of an ensemble under Levitt's pulse sequence with the field inhomogeneity. The "initial manifold" at $t_{\text{i}}=T/4$ has a spread in $\eta_{\text{i}}$ with a single $\phi_{\text{i}}$, while the "final manifold" at $t_{\text{f}}=T$ has a spread in $\phi_{\text{f}}$ with minimal spread in $\eta_{\text{f}}$, corresponding to refocusing in the $\eta$-direction.

\begin{figure}[ht!]
    \centering
    \includegraphics[width=1\linewidth]{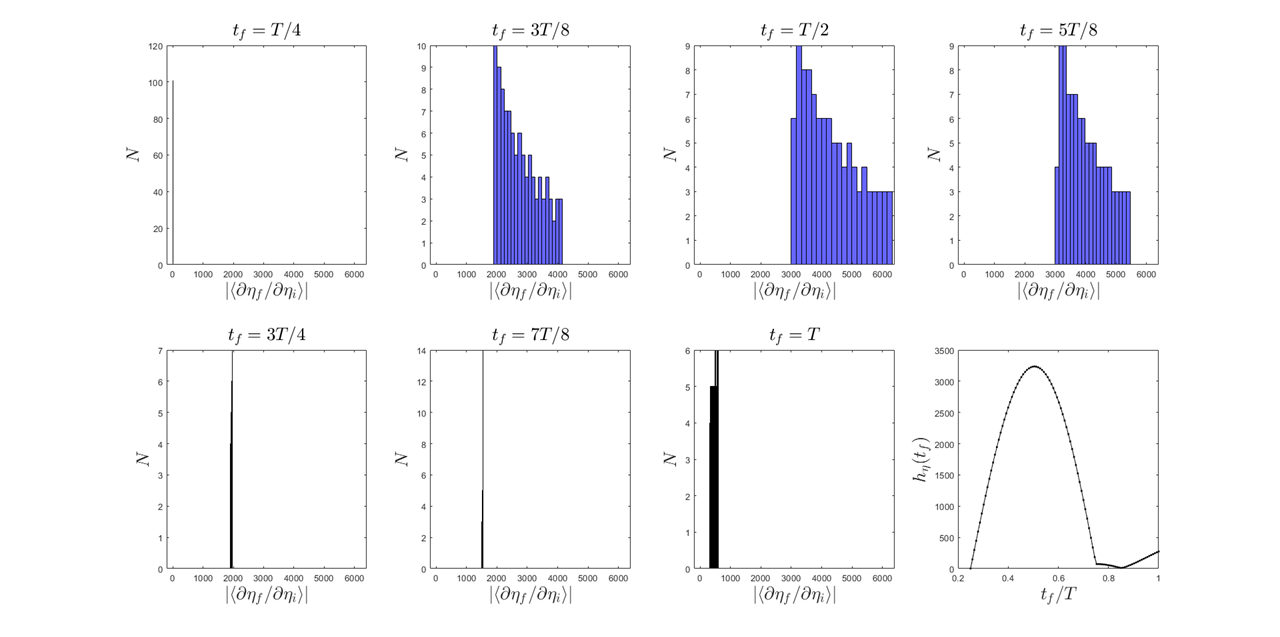}
    \caption{The time-evolution of a histogram representing the average stability element (in absolute value) $|\langle\partial\eta_{\text{f}}/\partial\eta_{\text{i}}\rangle|$ for the ensemble of field inhomogeneity, at seven moments in time. The refocusing in the $\eta$-direction is manifested by a collapsing of the histogram into a relatively narrow band at $t_{\text{f}}=T$. The rightmost subfigure in the lower panel shows the time-evolution of the range parameter $h_{\eta}$. The refocusing in the $\eta$-direction is clear.}
    \label{fig:stability_histogram_field_ih_eta}
\end{figure}

In the Rotating Frame, we launch at $t_{0}=0$ a central swarm of trajectories, from $\eta_{0}=1$, and a satellite swarm from $\eta_{0}-\Delta\eta_{0}=1-2\cdot10^{-6}$. From $t_{0}=0$ to $t_{\text{i}}=T/4$, the pulse keeps all trajectories in a line. For each moment $t_{\text{f}}\in[T/4,T]$, the set $\partial\eta_{\text{f}}/\partial\eta_{\text{i}}$ is calculated using finite-difference and binned according to eq. \ref{average-stability}. Figure  \ref{fig:stability_histogram_field_ih_eta} shows representative slices of this time-evolution: from $t_{\text{f}}=T/4$ to $t_{\text{f}}=T/2$, the histogram becomes wider, then it becomes narrower, until $t_{\text{f}}\approx7T/8$, where the range of the histogram becomes wider again, and eventually, at $t_{\text{f}}=T$ the histogram collapses to a relatively narrow column, compared to the maximal range at $t_{\text{f}}=T/2$. Note that between $t_{\text{f}}\approx3T/4$ and $t_{\text{f}}=1$ the refocusing is somewhat better, but does not take place in the vicinity of $\eta=-1$ (note that the histogram plots do not indicate where the refocusing takes place).

\begin{figure}[ht!]
    \centering
    \includegraphics[width=1\linewidth]{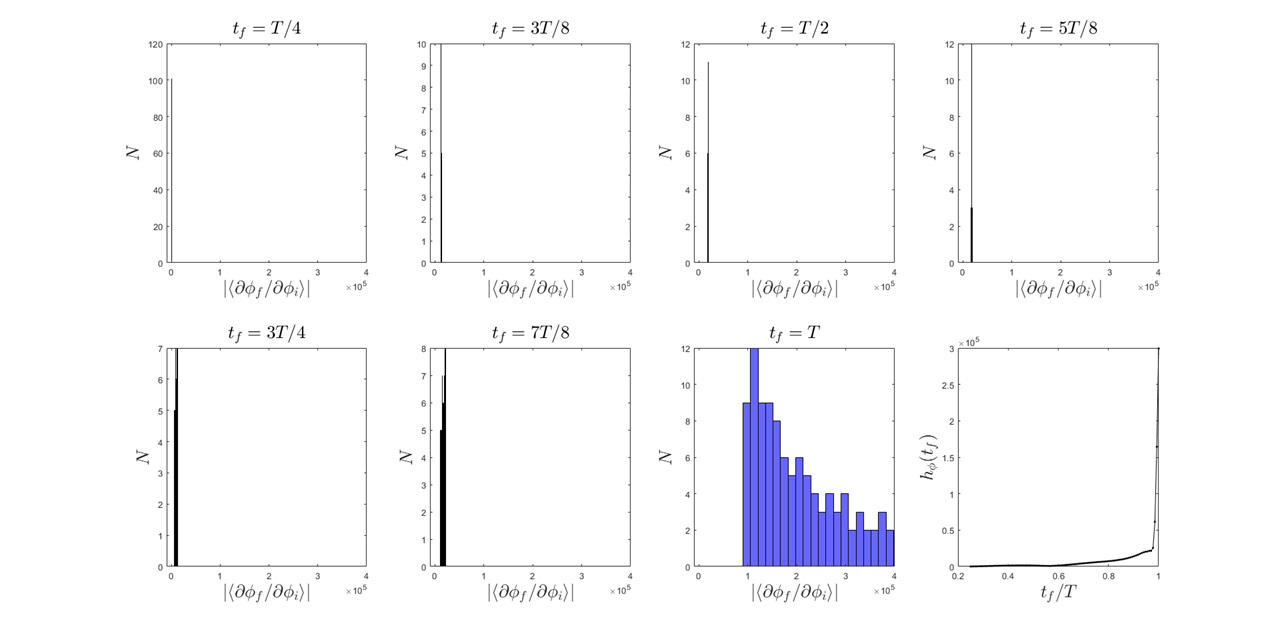}
    \caption{The time-evolution of a histogram representing the average stability element (in absolute value) $|\langle\partial\phi_{\text{f}}/\partial\phi_{\text{i}}\rangle|$ for the ensemble of field inhomogeneity, at seven moments in time. The anti-refocusing in the $\phi$-direction is manifested in a spreading of the histogram into a broad band at $t_{\text{f}}=T$. The rightmost subfigure in the lower panel shows the time-evolution of the range parameter $h_{\phi}$. The "anti-refocusing" in the $\phi$-direction is clear.}
    \label{fig:stability_histogram_field_ih_phi}
\end{figure}

For completeness, we repeat the same procedure in the $\phi$-direction. At $t_{0}=0$ we launch a central swarm of trajectories, from $\phi_{0}=0$  and a satellite swarm from $\phi_{0}+\Delta\phi_{0}=10^{-6}$ (and, to overcome the abovmentioned challenge, $\eta_{0}=1-10^{-6}$). For each moment $t_{\text{f}}\in[T/4,T]$, the set $\partial\phi_{\text{f}}/\partial\phi_{\text{i}}$ is calculated using finite-difference and binned according to eq. \ref{average-stability}. Figure \ref{fig:stability_histogram_field_ih_phi} shows representative slices of this time-evolution: from $t_{\text{f}}=T/4$ to $t_{\text{f}}=T$, the histogram becomes wider, and towards the end its range diverges in width to represent a kind of "anti-refocusing" in the $\phi$-direction.

\subsection{The ensemble of resonance offset}
Figure \ref{fig:Levitt_recovered} (right) shows the evolution of an ensemble under Levitt's pulse sequence with resonance offset. The "initial manifold" at $t_{\text{i}}=T/4$ has a spread in both $\phi_{\text{i}}$ and $\eta_{\text{i}}$, while the "final manifold" at $t_{\text{f}}=T$ comprises minimal spread in both $\phi_{\text{i}}$ and $\eta_{\text{i}}$, corresponding to refocusing in both the $\phi$- and $\eta$-directions.

\begin{figure}[ht!]
    \centering
    \includegraphics[width=1\linewidth]{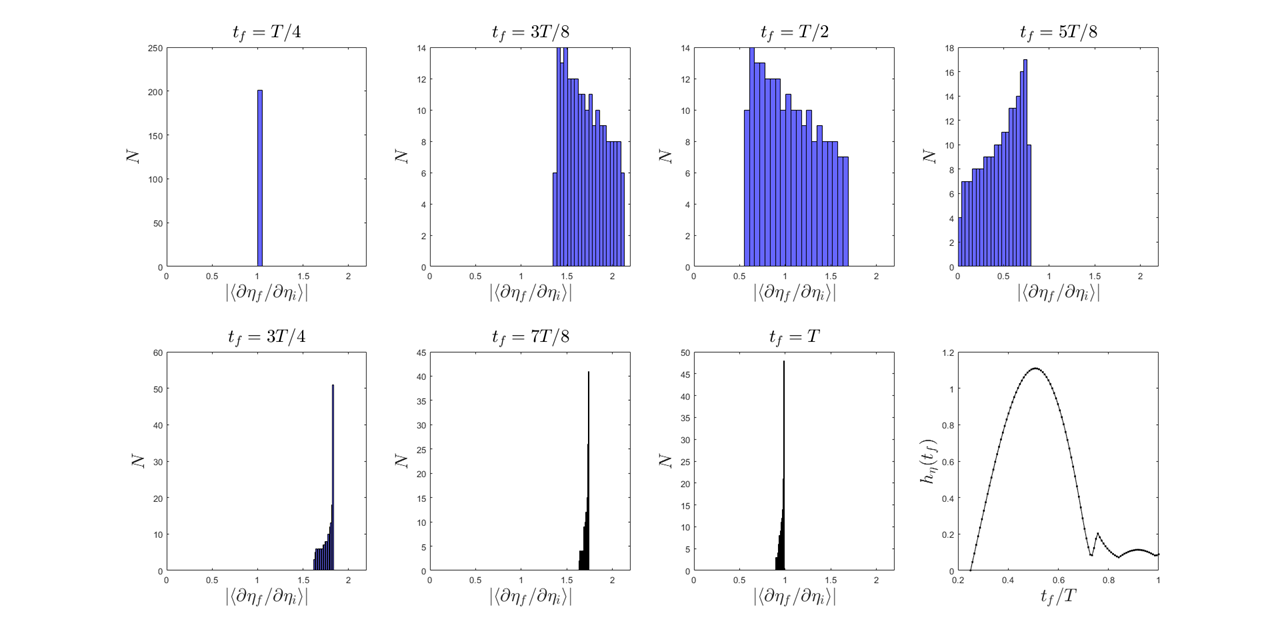}
    \caption{The time-evolution of a histogram presenting the average stability element (in absolute value) $|\langle\partial\eta_{\text{f}}/\partial\eta_{\text{i}}\rangle|$ for the ensemble of resonance offset, at seven moments in time. The refocusing in the $\eta$-direction is manifested by a collapsing of the histogram into a very narrow band at $t_{\text{f}}=T$. The rightmost subfigure in the lower panel shows the time-evolution of the range parameter $h_{\eta}$. The refocusing in the $\eta$-direction is clear.}
    \label{fig:stability_histogram_resonance_off_eta}
\end{figure}

\begin{figure}[ht!]
    \centering
    \includegraphics[width=1\linewidth]{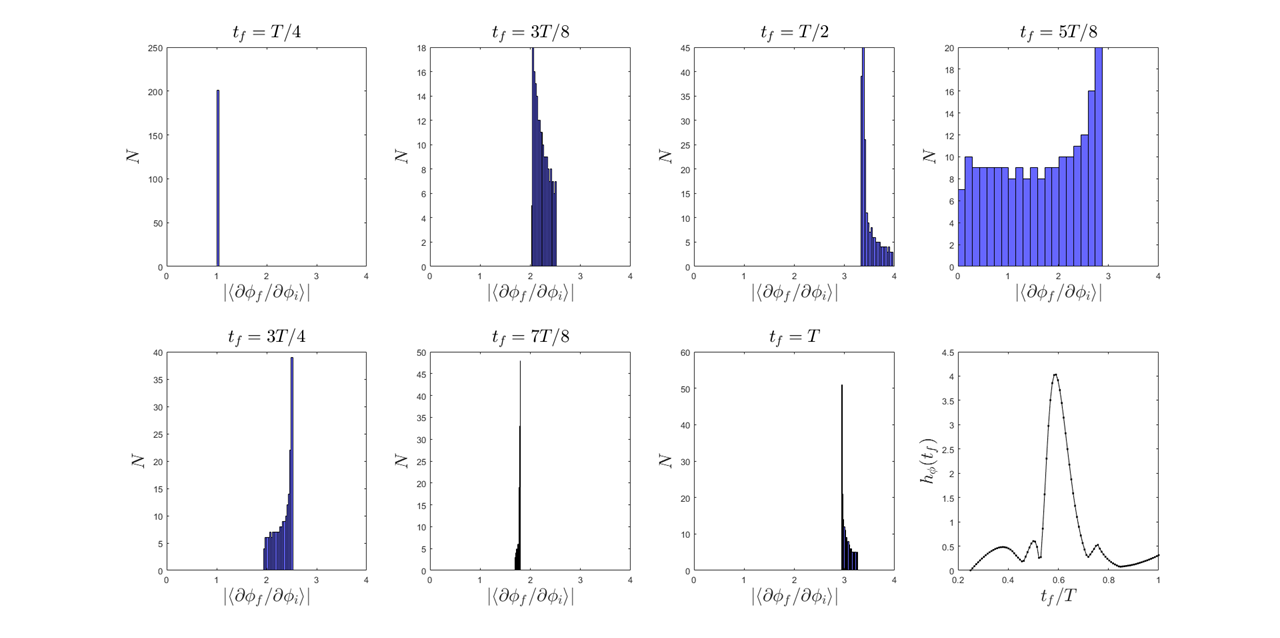}
    \caption{The time-evolution of a histogram presenting the average stability element (in absolute value) $|\langle\partial\phi_{\text{f}}/\partial\phi_{\text{i}}\rangle|$ for the ensemble of resonance offset, at seven moments in time. The eventual refocusing in the $\phi$-direction is manifested by a collapsing of the histogram into a very narrow band at $t_{\text{f}}=T$. The rightmost subfigure in the lower panel shows the time-evolution of the range parameter $h_{\phi}$. The refocusing in the $\phi$-direction is clear.}
    \label{fig:stability_histogram_resonance_off_phi}
\end{figure}

In the Rotating Frame, we propagate a central swarm and a satellite swarm with parameters as presented in the previous subsection. In contrast to the ensemble of field inhomogeneity, here, the trajectories are not aligned between $t_{0}=0$ and $t_{\text{i}}=T/4$, in any of the directions. For each moment $t_{\text{f}}\in[T/4,T]$, the sets $\partial\phi_{\text{f}}/\partial\phi_{\text{i}}$ and $ \partial\eta_{\text{f}}/\partial\eta_{\text{i}}$ are calculated using finite-difference and binned according to eq. \ref{average-stability}. Figures \ref{fig:stability_histogram_resonance_off_eta},\ref{fig:stability_histogram_resonance_off_phi} show representative slices of this time-evolution. At $t_{\text{f}}=T/4$, both histograms are wide, and they become even wider at later times until $t_{\text{f}}=T/2$ (for $\eta$) and $t_{\text{f}}\approx5T/8$ (for $\phi$). At this point, the ranges of both histograms become narrower and eventually, at $t_{\text{f}}=T$, the histograms collapse to narrow bands.

In the $\eta$-direction, the tightest refocusing occurs at $t_{\text{f}}=T$, although there are other moments where the refocusing is almost as good; in the $\phi$-direction, there is good refocusing at $t_{\text{f}}=T$ although there are regions around $t_{\text{f}}=0.4T$ and $t_{\text{f}}=0.8T$ where the refocusing is better. 

\subsection{Variation of the ensemble's width}
With the classical understanding in hand, we can address the question: why does the imperfection of field inhomogeneity conserve the width of the ensemble along the second and the third segment, while the imperfection of resonance offset expands or contracts the width along all segments?

For simplicity, we use a Euclidean measure of the width of an ensemble:
\begin{equation}
\label{width-1}
\sigma(t)=\sqrt{\frac{1}{N}\sum_{j=1}^{N}(\textbf{r}_{j}(t)-\bar{\textbf{r}}(t))^{2}},
\end{equation}
where $\textbf{r}_{j}(t)$ is a trajectory in the ensemble and $\bar{\textbf{r}}(t)$ is the mean of the ensemble; $N$ is the total number of trajectories in the ensemble. For convenience, we will suppress the time-parameterization below.

We differentiate $\sigma^{2}$ with respect to time to obtain:
\begin{equation}
\label{width-2}
\frac{d\sigma^{2}}{dt}=\frac{2}{N}\sum_{j=1}^{N}(\textbf{r}_{j}-\bar{\textbf{r}})^{T}\cdot(\dot{\textbf{r}}_{j}-\dot{\bar{\textbf{r}}}).
\end{equation}
We will now treat each of the four terms. The first term appears as quadratic form with a skew-symmetric operator:
\begin{equation}
S_{1,j}\equiv\ \textbf{r}_{j} \cdot \dot {\textbf{r}}_{j} =\textbf{r}_{j}^{T}\cdot \underline{\underline{\boldsymbol{\Omega}}}^{(j)}\cdot \textbf{r}_{j}=0.
\end{equation}
The second and third terms are cross terms:
\begin{equation}
S_{2,j}\equiv\textbf{r}_{j}^{T}\cdot\dot{\bar{\textbf{r}}}=\frac{1}{N}\textbf{r}_{j}^{T}\cdot\sum_{i}\underline{\underline{\boldsymbol{\Omega}}}^{(i)}\cdot\textbf{r}_{i}=\frac{1}{N}\sum_{i}\textbf{r}_{j}^{T}\cdot\underline{\underline{\boldsymbol{\Omega}}}^{(i)}\cdot\textbf{r}_{i},
\end{equation}
\begin{equation}
S_{3,j}\equiv \bar{\textbf{r}}^{T}\cdot\dot{\textbf{r}}_{j} =\frac{1}{N}\sum_{i}{\textbf{r}}_{i}^{T} \cdot\underline{\underline{\boldsymbol{\Omega}}}^{(j)}\cdot{\textbf{r}}_{j}=-\frac{1}{N}\sum_{i}{\textbf{r}}_{j}^{T} \cdot\underline{\underline{\boldsymbol{\Omega}}}^{(j)}\cdot{\textbf{r}}_{i},
\end{equation}
where in the last transition we used the anti-symmetric property of the triple product. Now we combine these terms, and as we recall the $j$-summation, we get:
\begin{equation}
 \sum_{j}S_{23,j}\equiv \sum_{j}S_{2,j}+S_{3,j}=\frac{1}{N}\sum_{i,j} {\textbf{r}}^{T}_{i}\cdot(\underline{\underline{\boldsymbol{\Omega}}}^{(j)}-\underline{\underline{\boldsymbol{\Omega}}}^{(i)})\cdot{\textbf{r}}_{j} =0,
 \end{equation}
 where again we used skew-symmetry. Finally, the fourth term leads to:
  \begin{equation}
 S_{4}\equiv \bar{\textbf{r}}^{T}\cdot\dot{\bar{\textbf{r}}}=\frac{1}{N^{2}} \sum_{i}{\textbf{r}}_{i}^{T}\cdot\sum_{k}\underline{\underline{\boldsymbol{\Omega}}}^{(k)}\cdot\textbf{r}_{k} =\frac{1}{N^{2}}\sum_{i,k} \textbf{r}_{i}^{T}\cdot\underline{\underline{\boldsymbol{\Omega}}}^{(k)}\cdot{\textbf{r}}_{k},
 \end{equation}
so we are left with the fourth term solely:
\begin{equation}
 \frac{d\sigma^{2}}{dt}=\frac{2}{N^{3}}\sum_{i,k} \textbf{r}_{i}^{T}\cdot\underline{\underline{\boldsymbol{\Omega}}}^{(k)}\cdot{\textbf{r}}_{k},
 \end{equation}
and it does not vanish in general.

In the case of field inhomogeneity, all rotations take place about fixed axes.
Numerical simulations indicate that along the first segment, $\sigma(t)$ varies, whereas along the second and the third segment, it is approximately constant. We will explain that.

Along the first segment we have:
\begin{equation}
\textbf{r}_{i}(t)=\begin{pmatrix}0\\ \sin \Omega_{1}^{(i)}t\\ \cos \Omega_{1}^{(i)}t \end{pmatrix}, \hspace{1cm}\underline{\underline{\boldsymbol{\Omega}}}^{(k)}=\begin{pmatrix}0 &&0 &&0 \\ 0  && 0 && \Omega_{1}^{(k)} \\ 0 && -\Omega_{1}^{(k)} && 0\end{pmatrix},\hspace{1cm} \underline{\underline{\boldsymbol{\Omega}}}^{(k)}\textbf{r}_{k}(t)=\Omega_{1}^{(k)}\begin{pmatrix}0 \\ \sin \Omega_{1}^{(k)}t \\ -\cos \Omega_{1}^{(k)}t\end{pmatrix},
\end{equation}
such that:
\begin{equation}
\frac{d\sigma^{2}}{dt}=-\frac{2}{N^{3}}\sum_{i,k}\Omega_{1}^{(k)}\cos((\Omega_{1}^{(i)}+\Omega_{1}^{(k)})t)\neq0.
\end{equation}
Note that we could factor out the $\Omega_{1}^{(k)}$ since all the members of the ensemble rotate about the same axis.
Along the second segment we have:
\begin{equation}
\textbf{{r}}_{i}(t)=\begin{pmatrix}z_{1}^{(i)}
\sin \Omega_{1}^{(i)}t\\ y_{1}^{(i)}\\ z_{1}^{(i)}\cos \Omega_{1}^{(i)}t \end{pmatrix}, \hspace{1cm}\underline{\underline{\boldsymbol{\Omega}}}^{(k)}=\begin{pmatrix}0 &&0 &&-\Omega_{1}^{(k)} \\ 0  && 0 && 0 \\ \Omega_{1}^{(k)} && 0 && 0\end{pmatrix},\hspace{1cm}\underline{\underline{\boldsymbol{\Omega}}}^{(k)}\textbf{r}_{k}(t)=\Omega_{1}^{(k)}\begin{pmatrix}-z_{1}^{(k)}\cos \Omega_{1}^{(k)}t \\ 0 \\ z_{1}^{(k)}\sin \Omega_{1}^{(k)}t\end{pmatrix},
\end{equation}
such that, approximately:
\begin{equation}
\frac{d\sigma^{2}}{dt}=-\frac{2}{N^{3}}\sum_{i,k}\Omega_{1}^{(k)}z_{1}^{(i)}z_{1}^{(k)}\sin((\Omega_{1}^{(i)}+\Omega_{1}^{(k)})t)\approx-\frac{2}{N^{3}}\sum_{i,k}\Omega_{1}^{(k)}z_{1}^{(0)2}\sin((\Omega_{1}^{(i)}+\Omega_{1}^{(k)})t)=0.
\end{equation}
The approximation is justified as we consider a narrow band of $\Omega_{1}^{(i)}$ around the nominal frequency, and:
\begin{equation}
z_{1}^{(0)}=\cos \Omega_{1}^{(0)}T/4=\cos \pi/2=0.
\end{equation}
Now we will examine the third segment, where we have:
\begin{equation}
\textbf{{r}}_{i}(t)=\begin{pmatrix}x_{2}^{(i)}
\\\ y_{2}^{(i)}\cos \Omega_{1}^{(i)}t -z_{2}^{(i)}\sin \Omega_{1}^{(i)}t\\ y_{2}^{(i)}\sin \Omega_{1}^{(i)}t +z_{2}^{(i)}\cos \Omega_{1}^{(i)}t \end{pmatrix}, \hspace{1cm} \underline{\underline{\boldsymbol{\Omega}}}^{(k)}\textbf{r}_{k}(t)=\Omega_{1}^{(k)}\begin{pmatrix}0
\\\ y_{2}^{(k)}\sin \Omega_{1}^{(k)}t +z_{2}^{(k)}\cos \Omega_{1}^{(k)}t\\ -y_{2}^{(k)}\cos \Omega_{1}^{(k)}t -z_{2}^{(k)}\sin \Omega_{1}^{(k)}t \end{pmatrix},
\end{equation}
such that, in the same sense as the previous segment, we get:
\begin{equation}
\frac{d\sigma^{2}}{dt}\approx-\frac{2}{N^{3}}\sum_{i,k}\Omega_{1}^{(k)}y_{2}^{(0)2}\sin((\Omega_{1}^{(i)}-\Omega_{1}^{(k)})t)\approx0,
\end{equation}
where
\begin{equation}
y_{2}^{(0)}=y_{1}^{(0)}\sin \Omega_{1}^{(0)}T/2=\sin \pi/2=1,\hspace{1cm} z_{2}^{(0)}=z_{1}^{(0)}\cos \Omega_{1}^{(0)}T/2=0,
\end{equation}
and the approximation is justified as we consider a narrow band of $\Omega_{1}^{(i)}$ around the nominal frequency, such that:
\begin{equation}
\sin((\Omega_{1}^{(i)}-\Omega_{1}^{(k)})t)\approx\sin(0)=0.
\end{equation}
In the case of resonance offset, each trajectory rotates about a different axis which is inclined with respect to the $xy$ plane. For example, in along the first segment, the rotation is specified by the matrix
\begin{equation}
\underline{\underline{\boldsymbol{\Omega}}}^{(k)}=\begin{pmatrix}0 &&-\Delta^{(k)} &&0 \\ \Delta^{(k)} && 0 && {\Omega}_{1} \\ 0 && -{\Omega}_{1} && 0\end{pmatrix},\end{equation}
and a time-dependent trajectory $\textbf{r}_{i}(t)$ outcomes from the exponentiation of this matrix, using Rodrigues' formula; in the former case, the exponentiation was trivial and led to $x$- or $y$- rotation matrices. Numerical simulations indicate that along all segments, $\sigma(t)$ varies. We will explain that based on the previous derivations.

In contrast to the former case, we cannot factor out either of the frequency components, such that each sum over $i,k$ is supposed to contain various combinations of trigonometric functions, rather than a single function as before. The prefactors of the summands will contain the initial conditions of each segment, which in general do not vanish, as opposed to the previous case. Altogether, $d\sigma^{2}/dt$ is not predicted to vanish at any segment.

\begin{figure}[ht!]
    \centering
    \includegraphics[width=0.7\linewidth]{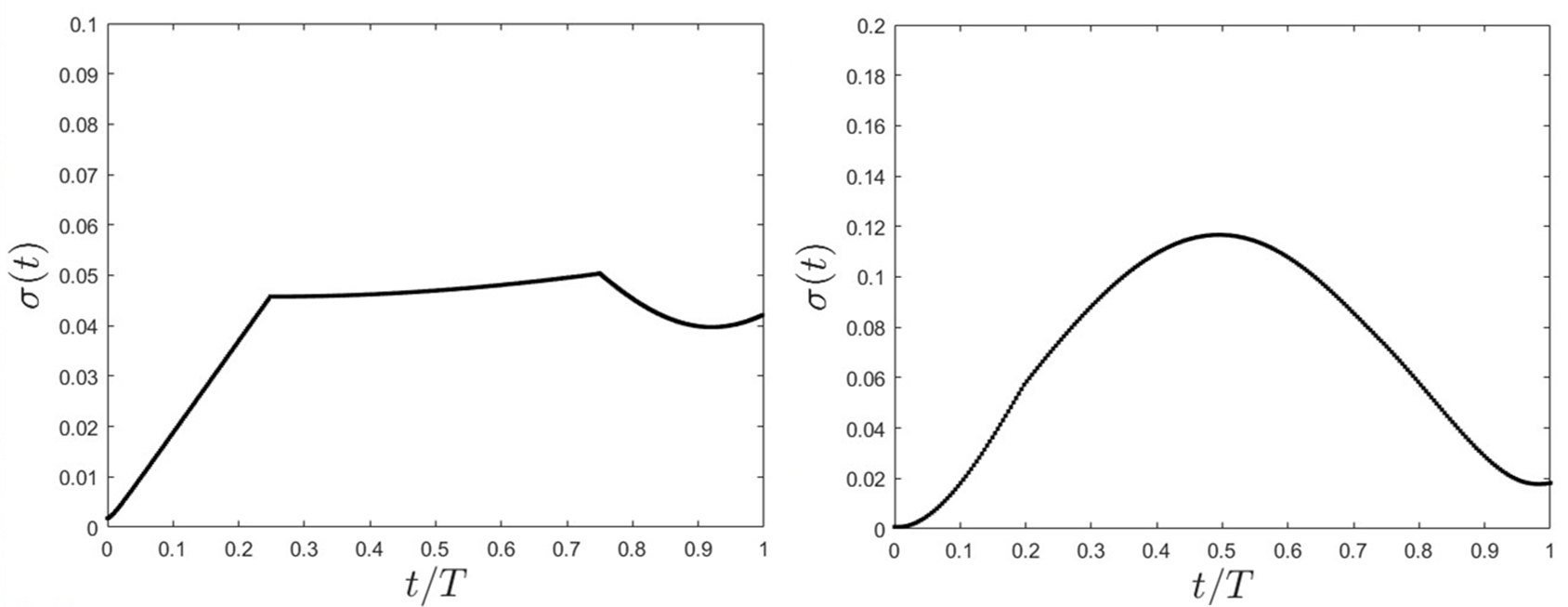}
    \caption{A time-evolution of the width $\sigma(t)$ in the case of field inhomogeneity (left) and in the case of resonance offsets (right) for Levitt's pulse sequence. It is evident that the width is approximately conserved along the second and the third segments in the former case, and definitely not conserved in the latter case.}
    \label{fig:width1}
\end{figure}

Figure \ref{fig:width1} demonstrates the numerical results for the Levitt's sequence. The left subfigure corresponds to field inhomogeneity and the right subfigure corresponds to resonance offsets. Note that the width of the ensemble is preserved in second and the third segment of the former case, and not conserved in the latter case.

\subsection{Conclusion}
In this work, we suggest another justification based on the classical mechanical concept of a stability matrix. The motion on the Bloch Sphere is mapped to a canonical system of coordinates and the focusing of an ensemble corresponds to caustics, or the vanishing of an appropriate stability matrix element in the canonical coordinates. To the best of our knowledge, this is the first work that introduces a canonical version of the Bloch Equations and furthermore, that applies the concepts of classical stability analysis and caustics to the investigation of dynamics on the Bloch Sphere.

The caustics in CPs are unusual in the sense that each member is governed by a slightly different Hamiltonian, such that averaging is required for the stability analysis. The average stability matrix elements turned out to be particularly informative: when they collapse to a narrow band, the ensemble refocuses, and when they diverge, the ensemble exhibits a maximal spread. This is a clear, visual, and quantitative way to track robustness.

Our approach highlights the directionality of the refocusing of the ensemble on the Bloch Sphere, revealing how different ensembles refocus along different directions. As a case study, we investigated the $90(x)180(y)90(x)$ CP introduced by Levitt, where the approach provides a new perspective into why this CP is effective: the focusing produced by Levitt's CP corresponds to a caustic, as manifested in the elements of the stability matrix. Levitt's perturbative treatment was seen to correspond to one element of the classical stability matrix.

The approach clarifies why the $90(x)180(y)90(x)$ CP changes the width of the ensemble in the case of field inhomogeneity, as opposed to simply a rotation of the axes in the case of resonance offset. In the case of field inhomogeneity, the pulse leads to refocusing in the $\eta$-direction, while in the $\phi$-direction the ensemble diverges, meaning the refocusing is highly directional. In the case of resonance offsets, refocusing occurs in both directions. This directional aspect goes beyond what was discussed by Levitt. 

Although we focused here on the $90(x)180(y)90(x)$ pulse sequence, our method can be applied to any pulse sequence. It does not rely on a specific form of the pulse or any assumption about the ensemble; all one needs is the pulse sequence and its imperfections. In this sense, we believe that our classical point of view can be useful for analyzing almost all types of CPs.

In summary, we believe that the transfer of ideas from classical mechanics to quantum control is intellectually satisfying, and that notions like stability analysis, manifolds and caustics, can provide new insights into this widely studied system. This ultimately may suggest new pulse sequences for NMR, optical spectroscopy, quantum information processing and quantum computing. J.B. and D.J.T. thank Jacob Higer and Ilya Kuprov for helpful discussions.

\bibliographystyle{apsrev4-2}
\bibliography{main.bbl}

\end{document}